# The ruggedness of protein-protein energy landscape and the cutoff for 1/$r^n$ potentials

**Anatoly M. Ruvinsky[1] and Ilya A. Vakser[2]**

[1]Anatoly M. Ruvinsky, Center for Bioinformatics, The University of Kansas, 2030 Becker Drive, Lawrence, Kansas 66047; tel: (785) 864-1057, fax: (785) 864-5558, email: ruvinsky@ku.edu

[2]Center for Bioinformatics and Department of Molecular Biosciences, The University of Kansas, Lawrence, Kansas 66047; 2030 Becker Drive, Lawrence, Kansas 66047; tel: (785) 864-1057, fax: (785) 864-5558, email: vakser@ku.edu

## ABSTRACT

The interaction cutoff contribution to the ruggedness of protein-protein energy landscape (the artificial ruggedness) is studied in terms of relative energy fluctuations for $1/r^n$ potentials based on a simplistic model of a protein complex. Contradicting the principle of minimal frustration, the artificial ruggedness exists for short cutoffs and gradually disappears with the cutoff increase. The critical values of the cutoff were calculated for each of eleven popular power-type potentials with $n=0\div9$, 12 and for two thresholds of 5% and 10%. The artificial ruggedness decreases to tolerable thresholds for cutoffs longer than the critical ones. The results showed that for both thresholds the critical cutoff is a non-monotonic function of the potential power $n$. The functions reach the maximum at $n=3\div4$ and then decrease with the increase of the potential power. The difference between two cutoffs for 5% and 10% artificial ruggedness becomes negligible for potentials decreasing faster than $1/r^{12}$. The results suggest that cutoffs longer than critical ones can be recommended for protein-protein potentials.

# INTRODUCTION

Protein binding can be explained in terms of the funnel-based concept initially developed to describe protein folding [1-10]. The concept suggests that unbound proteins are guided by the slope of the rugged energy landscape funnel into the bound state. The nature of the ruggedness and related effects is a subject of active research [11-14]. Highly frustrated interactions are observed on the protein surface near the binding site [14]. Mechanical unfolding experiments to measure the scale of the landscape ruggedness of proteins and RNAs have been suggested [15] and performed [16].

The amplitude of the of protein-protein energy landscape ruggedness has a component associated with the range of the energy potentials [12]. The range of non-bonded inter-atomic interactions and related truncation methods are known to play an important role in protein folding [17-22], protein-protein docking [4,12], and all atom molecular dynamics and Monte-Carlo simulations of macromolecules and liquids [23-29]. The cutoff is one of only two parameters used in coarse-grained normal mode analysis and elastic networks of proteins and their assemblies [30,31]. The choice of the cutoff affects the functional form and performance of knowledge-based potentials in small molecule docking [32]. The importance of long range interactions for protein stability [33], protein folding [34] and RNA binding [35] has been revealed experimentally.

The interactions are usually truncated at specific cutoff distances to reduce a number of interacting pairs of atoms or atomic groups in order to make feasible large scale macromolecular calculations. Despite the considerable progress achieved in methodology and computer power, the cutoff-related artifacts are still a bottleneck in macromolecular modeling. In comparison with other modeling approaches, the protein docking community has been less focused on the problem. Our attention to the cutoff problem is motivated by observations that the choice of larger cutoffs results in the ruggedness depression [12] and thus in smooth protein-protein energy landscapes [4,36], which according with the principal of minimal frustration, [1,7,37] better approximate the actual binding landscape. Similar effects of the energy landscape smoothing due to the cutoff extension have been found in studies of liquids and atomic clusters [38-43], helix dimers [4,44,45] and protein complexes [4,36,46].

We have recently demonstrated [12,47] that short cutoffs perturb protein-protein energy landscape and thus lead to false minima, changed positions and altered shape of true conformation-based minima. Such changes of the landscape impede the search for the global minimum in protein docking [4] and introduce errors in calculations of binding free energy [6,48-50]. The false minima cause the artificial ruggedness of the energy landscape. The fine structure of the funnel or conformational substates [51] can be blurred due to the artificial ruggedness. The amplitude of the artificial ruggedness decreases with the increase of the cutoff [12]. Thus it is important to know the cutoffs for different potentials that form a minimally frustrated funnel-like landscape, while allowing extensive calculations. The low boundary of the optimal range, called further the critical cutoff, corresponds to a tolerable frustration of the energy landscape.

In this paper, we focus on determination of critical cutoffs for eleven power-type potentials at two thresholds of 5% and 10% of the artificial ruggedness. For cutoffs longer than the critical ones, the artificial ruggedness drops below these thresholds. We analyze dependence of the critical cutoffs on the potential power $n$ and discuss practical implications of the results for protein docking and protein folding.

**THEORY**

Using a simplistic model of a sandwich-like protein complex [12,47] (see also Ref [52]), we can write an energy of the complex as

$$E = 2\pi\rho^2 S \int_a^R dx \int_x^R \varepsilon(r)(r^2 - rx)dx, \quad (1)$$

where $S$ is the average interface area, $\rho$ is the protein density, $\varepsilon(r) \sim r^{-n}$ is the power-type inter-atomic potential, $R$ is the interaction cutoff, $a$ is the minimal distance between two proteins (see Ref [12] for the detail derivation of Eq 1). For simplicity, we assume that potential $\varepsilon(r)$ does not depend on atom type. If we slightly enlarge the distance between the two proteins, the energy of the new conformation will be

$$E_c = 2\pi\rho^2 S_n \int_{a+\delta a}^R dx \int_x^R \varepsilon(r)(r^2 - rx)dr, \quad (2)$$

where $\delta a$ is the relative shift of proteins, $S_c$ is the new interface. The artificial ruggedness of the landscape manifests itself as the deviation of the relative energy change $\delta E(R)/E(R) = (E_c(R) - E(R))/E(R)$ from its value at the asymptotically large cutoff $R_L$

$$v_{art}(R) = \left( \frac{E_c(R) - E(R)}{E(R)} - \frac{E_c(R_L) - E(R_L)}{E(R_L)} \right) \times 100\% \quad (3)$$

Intermolecular energy can be written as a sum of pair-wise inter-atomic interactions described by model potentials and direct electrostatic potentials [53]. As a rule, these potentials (e.g. Lennard-Jones potentials and their modifications, multipole-multipole potentials) have a form of an expansion over $r^{-n}$. The simple form of the power potentials allows analytical evaluation of Eq 3 for all integer $n$. Further, assuming $S_c \approx S$, $a = 2.8\text{Å}$, $\delta a = 0.5\text{Å}$ and $R_L = 30\text{Å}$, we compute two critical cutoffs for each of eleven potentials with $n = 0 \div 9, 12$ under the condition that the artificial ruggedness drops below the threshold of 10% or 5% for cutoffs longer than the critical ones.

**RESULTS AND DISCUSSIONS**

Summarized in Table I are the results of the calculations of the asymptotical behavior of the relative energy change at large cutoffs, of the artificial ruggedness and the critical cutoffs for different power-type potentials. The relative energy change $\delta E(R)/E(R)$ asymptotically approaches zero for $n \leq 4$, and approaches a constant $-(n-4)\delta a/a$ for $n>4$. The artificial ruggedness is a decreasing function of the cutoff for each of eleven potentials. The results show that both critical cutoffs depend non-monotonically on the potential power $n$ (Fig. 1). They increase up to the maximum at $n=4$ and then decrease with the power increase. The estimates of the critical cutoff for $n=6$ and 12 are in a good agreement with our previously published results based on use of a soft Lennard-Jones potential on a set of 66 protein complexes [12]. The difference between two cutoffs, which correspond to artificial ruggedness of 10% and 5%, decreases for $n>6$, and becomes negligible for potentials decreasing faster than $1/r^{12}$.

Since protein folding and protein binding are similar processes in terms of the landscape characteristics, including the funnel concept, we may expect that our results have implications to protein folding. Highly systematic attempts have been undertaken to design pair potentials for protein folding.[54-57] Using machine learning algorithms, the authors of these studies clearly showed that a set of contact potentials with cutoffs of 8.5 Å or 9Å, which guarantees the native structure energies lower than those of the decoys, does not exist. Then, using different resolutions of the potential functions, the same learning algorithm, and the 9Å cutoff, the flexible functional forms of potentials were optimized. Based on the performance of the potentials, it was noted that it is impossible to find a pair potential with that flexible form that recognizes all native folds (Ref. 55,56). The choice of the cutoff may partly explain these results and thus encourage new attempts to parameterize potentials for longer ranges. Indeed, the range of 9Å is less than the critical cutoffs of power potentials for $n\leq6$ and the artificial ruggedness threshold of 5%, or for $n\leq8$ and the artificial ruggedness threshold of 10% (see Table I). For example, the artificial ruggedness of the energy landscape described by contact or Coulomb potentials cutoff at 8-9 Å is ~17-19%. Since substantially frustrated landscapes are not adequate approximations of actual energy profiles due to the principle of minimal frustration, [1-3,7,58] the above studies [58] had limited chances to detect the actual parameters of the interactions. Our results suggest that using longer cutoffs with such algorithms may improve the potentials.

**CONCLUSIONS**

Studies of ruggedness of protein-protein energy landscape are important for understanding the connection between protein structure, function, and dynamics. We have analyzed energy fluctuations and the artificial contribution to the ruggedness of the protein-protein energy landscape by limited range interactions described by $1/r^n$ potentials. The results show that the undesirable artificial ruggedness exists for short cutoffs and gradually disappears with the cutoff increase. We calculated the critical values of the cutoff for each of eleven popular power-type potentials with $n=0\div9, 12$ and for two thresholds of 5% and 10%. We demonstrated that for both thresholds, the critical cutoff is a non-monotonic function of the potential power $n$. These functions reach the maximum at $n=3\div4$ and then decrease with the increase of the potential power. The difference between the cutoffs for 5% and 10% artificial ruggedness becomes negligible for potentials decreasing faster than $1/r^{12}$. The results suggest that the cutoffs longer than the critical ones can be recommended for protein-protein potentials.

**ACKNOWLEDGMENT**

The study was supported by R01 GM074255 grant from NIH.

**LEGEND TO FIGURES**

**Figure 1:** The critical cutoff as a function of the potential power.

# REFERENCES

1. Bryngelson JD, Onuchic JN, Socci ND, Wolynes PG. Funnels, pathways, and the energy landscape of protein folding: A synthesis. Proteins 1995;21:167-195.

2. Dill KA. Polymer principles and protein folding. Protein Sci 1999;8:1166-1180.

3. Tsai C-J, Kumar S, Ma B, Nussinov R. Folding funnels, binding funnels, and protein function. Protein Sci 1999;8:1181-1190.

4. Vakser IA. Long-distance potentials: An approach to the multiple-minima problem in ligand-receptor interaction. Protein Eng 1996;9:37-41.

5. Tovchigrechko A, Vakser IA. How common is the funnel-like energy landscape in protein-protein interactions? Protein Sci 2001;10:1572-1583.

6. Minh DDL, Bui JM, Chang CE, Jain T, Swanson JMJ, McCammon JA. The entropic cost of protein-protein association: A case study on acetylcholinesterase binding to fasciculin-2. Biophys J: Biophys Lett 2005;89:L25-L27.

7. Wolynes PG. Recent successes of the energy landscape theory of protein folding and function. Quart Rev Biophys 2005;38:405-410.

8. Camacho CJ, Vajda S. Protein docking along smooth association pathways. Proc Natl Acad Sci USA 2001;98(19):10636-10641.

9. Camacho CJ, Weng Z, Vajda S, DeLisi C. Free energy landscapes of encounter complexes in protein-protein association. Biophys J 1999;76:1166-1178.

10. Hunjan J, Tovchigrechko A, Gao Y, Vakser IA. The size of the intermolecular energy funnel in protein-protein interactions. Proteins 2008:Epub Jan 23.

11. O'Toole N, Vakser IA. Large-scale characteristics of the energy landscape in protein-protein interactions. Proteins 2008;71:144-152.

12. Ruvinsky AM, Vakser IA. Interaction cutoff effect on ruggedness of protein-protein energy landscape. Proteins 2008;70:1498-1505.

13. Sutto L, Latzer J, Hegler JA, Ferreiro DU, Wolynes PG. Consequences of localized frustration for the folding mechanism of the IM7 protein. Proc Natl Acad Sci USA 2007;104:19825–19830.

14. Ferreiro DU, Hegler JA, Komives EA, Wolynes PG. Localizing frustration in native proteins and protein assemblies. Proc Natl Acad Sci USA 2007;104:19819–19824.

15. Hyeon C, Thirumalai D. Can energy landscape roughness of proteins and RNA be measured by using mechanical unfolding experiments? Proc Natl Acad Sci USA 2003;100:10249-10253.

16. Nevo R, Brumfeld V, Kapon R, Hinterdorfer P, Reich Z. Direct measurement of protein energy landscape roughness. EMBO Rep 2005;6:482-486.

17. Go N, Taketomi H. Respective roles of short- and long-range interactions in protein folding. Proc Natl Acad Sci USA 1978;75:559-563.

18. Govindarajan S, Goldstein RA. Optimal local propensities for model proteins. Proteins 1995;22:413-418.

19. Abkevich VI, Gutin AM, Shakhnovich EI. Impact of local and non-local interactions on thermodynamics and kinetics of protein folding. J Mol Biol 1995;252:460-471.

20. Doyle R, Simons K, Qian H, Baker D. Local interactions and the optimization of protein folding. Proteins 1997;29:282-291.

21. Gromiha MM, Selvaraj S. Importance of long-range interactions in protein folding. Biophys Chem 1999;77:49-68.

22. Faisca PF, Telo da Gama MM, Nunes A. The Go model revisited: Native structure and the geometric coupling between local and long-range contacts. Proteins 2005;60:712-722.

23. Brooks CL, Pettitt BM, Karplus M. Structural and energetic effects of truncating long ranged interactions in ionic and polar fluids. J Chem Phys 1985;83:5897-5908.

24. Harvey SC. Treatment of electrostatic effects in macromolecular modeling. Proteins 1989;5:78-92.

25. Loncharich RJ, Brooks BR. The effects of truncating long-range forces on protein dynamics. Proteins 1989;6:32-45.

26. Gilson MK. Theory of electrostatic interactions in macromolecules. Curr Opin Struct Biol 1995;5:216-223.

27. Haluk R, McCammon JA. Correcting for electrostatic cutoffs in free energy simulations: Toward consistency between simulations with different cutoffs. J Chem Phys 1998;108:9617-9623.

28. Norberg J, Nilsson L. On the truncation of long-range electrostatic interactions in DNA. Biophys J 2000;79:1537-1553.

29. Sagui C, Darden TA. Molecular dynamics simulations of biomolecules: Long-range electrostatic effects. Ann Rev Biophys Biomol Struct 1999;28:155-179.

30. Bahar I, Atilgan AR, Erman B. Direct evaluation of thermal fluctuations in proteins using a single-parameter harmonic potential. Folding & Design 1997;2:173-181.

31. Hinsen K. Analysis of domain motions by approximate normal mode calculations. Proteins 1998;1998:417-429.

32. Ruvinsky AM, Kozintsev AV. The key role of atom types, reference states, and interaction cutoff radii in the knowledge-based method: New variational approach. Proteins 2005;58:845-851.

33. Grimsley GR, Shaw KL, Fee LR, Alston RW, Huyghues-Despointes BM, Thurlkill RL, Scholtz JM, Pace CN. Increasing protein stability by altering long-range Coulombic interactions. Protein Sci 1999;8:1843-1849.

34. Klein-Seetharaman J, Oikawa M, Grimshaw SB, Wirmer J, Duchardt E, Ueda T, Imoto T, Smith LJ, Dobson CM, Schwalbe H. Long-range interactions within a nonnative protein. Science 2002;295:1719-1722.

35. Lafuente E, Ramos R, Martinez-Salas E. Long-range RNA-RNA interactions between distant regions of the hepatitis C virus internal ribosome entry site element. J Gen Virol 2002;83:1113-1121.

36. Vakser IA, Matar OG, Lam CF. A systematic study of low-resolution recognition in protein-protein complexes. Proc Natl Acad Sci USA 1999;96:8477-8482.

37. Bryngelson JD, Wolynes PG. Intermediates and barrier crossing in a random energy model (with applications to protein folding). J Phys Chem 1989;93:6902-6915.

38. Stillinger FH, Stillinger DK. Cluster optimization simplified by interaction modification. J Chem Phys 1990;93:6106-6107.

39. Braier PA, Berry RS, Wales DJ. How the range of pair interactions governs features of multidimensional potentials. J Chem Phys 1990;93:8745-8756.

40. Doye JPK, Wales DJ. The effect of the range of the potential on the structure and stability of simple liquids: From clusters to bulk, from sodium to C60. J Phys B: At Mol Opt Phys 1996;29:4859–4894.

41. Miller MA, Doye JPK, Wales DJ. Structural relaxation in Morse clusters: Energy landscapes. J Chem Phys 1999;110:328-334.

42. Whitfield TW, Straub JE. Gravitational smoothing as a global optimization strategy. J Comput Chem 2002;23(11):1100-1102.

43. Wawak RJ, Wimmer MM, Scheraga HA. Application of the diffusion equation method of global optimization to water clusters. J Phys Chem 1992;96:5138-5145.

44. Vakser IA, Jiang S. Strategies for modeling the interactions of the transmembrane helices of G-protein coupled receptors by geometric complementarity using the GRAMM computer algorithm. Methods Enzym 2002;343:313-328.

45. Pappu RV, Marshall GR, Ponder JW. A potential smoothing algorithm accurately predicts transmembrane helix packing. Nature Struct Biol 1999;6(1):50-55.

46. Tovchigrechko A, Wells CA, Vakser IA. Docking of protein models. Protein Sci 2002;11:1888-1896.

47. Ruvinsky AM, Vakser IA. Chasing funnels on protein-protein energy landscapes at different resolutions. Biophys J 2008: doi:10.1529/biophysj.108.132977.

48. Ruvinsky AM, Kozintsev AV. New and fast statistical-thermodynamic method for computation of protein-ligand binding entropy substantially improves docking accuracy. J Comput Chem 2005;26:1089-1095. Ruvinsky AM. Role of binding entropy in the refinement of protein-ligand docking predictions: Analysis based on the use of 11 scoring functions. J Comput Chem 2007;28:1364-1372.

49. Alsallaq R, Zhou HX. Prediction of protein-protein association rates from a transition-state theory. Structure 2007;15:215–224.

50. Ruvinsky AM. Calculations of protein-ligand binding entropy of relative and overall molecular motions. J Comput Aided Mol Des 2007;21:361-370.

51. Frauenfelder H, McMahon BH, Austin RH, Chu K, Groves JT. The role of structure, energy landscape, dynamics, and allostery in the enzymatic function of myoglobin. Proc Natl Acad Sci USA 2001;98:2370-2374.

52. Lukatsky DB, Zeldovich KB, Shakhnovich EI. Statistically enhanced self-attraction of random patterns. Phys Rev Lett 2006;97:178101.

53. Kaplan IG. Intermolecular Interactions: Physical Picture, Computational Methods and Model Potentials. Chichester: John Wiley & Sons, Ltd; 2006.

54. Vendruscolo M, Domany E. Pairwise contact potentials are unsuitable for protein folding. J Chem Phys 1998;109:11101-11108.

55. Tobi D, Elber R. Distance-dependent, pair potential for protein folding: Results from linear optimization. Proteins 2000;41:40-46.

56. Tobi D, Shafran G, Linial N, Elber R. On the design and analysis of protein folding potentials. Proteins 2000;40:71-85.

57. Vendruscolo M, Najmanovich R, Domany E. Protein folding in contact map space. Phys Rev Lett 1999;82:656-659.

58. Miller DW, Dill KA. Ligand binding to proteins: The binding landscape model. Protein Sci 1997;6:2166-2179.

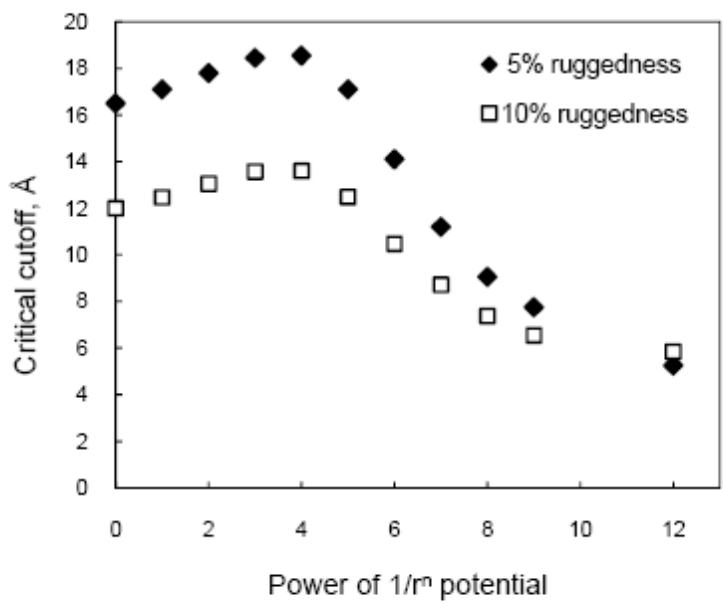

**Figure 1:** The critical cutoff as a function of the potential power.

**Table I. Asymptotical behavior of the relative energy change for different potentials**

| Types of interatomic interactions | The relative energy change $\delta E(R)/E(R)$ | Asymptotical behavior of $\delta E(R)/E(R)$ at large cutoffs | Critical cutoffs for $v_{art}(R)$ lower than 10% and 5%, Å |
|---|---|---|---|
| $1/r^0 = const$ | $\frac{\delta a}{R}\left(-2+3\frac{a}{R}-\frac{a^3}{R^3}\right)\left(\frac{3}{4}-2\frac{a}{R}+\frac{3a^2}{2R^2}-\frac{a^4}{4R^4}\right)^{-1}$ | $-\frac{8\delta a}{3R}$ | 12 / 16.5 |
| $1/r$ | $-\frac{3\delta a}{R-a}$ | $-\frac{3\delta a}{R}$ | 12.5 / 17.1 |
| $1/r^2$ | $-\frac{\delta a}{R}\left(1-\frac{a}{R}+\frac{a}{R}\ln\frac{a}{R}\right)\left(\frac{1}{4}-\frac{a}{R}+\frac{3a^2}{4R^2}-\frac{a^2}{2R^2}\ln\frac{a}{R}\right)^{-1}$ | $-\frac{4\delta a}{R}$ | 13.1 / 17.8 |
| $1/r^3$ | $-\frac{\delta a}{R}\left(-1+\frac{a}{R}-\ln\frac{a}{R}\right)\left(\frac{1}{2}-\frac{a^2}{2R^2}+\frac{a}{R}\ln\frac{a}{R}\right)^{-1}$ | $-2\frac{\delta a}{R}\ln\frac{R}{a}$ | 13.6 / 18.5 |
| $1/r^4$ | $-\frac{\delta a}{R}\left(-1+\frac{R}{2a}+\frac{a}{2R}\right)\left(-\frac{3}{4}+\frac{a}{R}-\frac{a^2}{4R^2}-\frac{1}{2}\ln\frac{a}{R}\right)^{-1}$ | $-\frac{\delta a}{a\ln(R/a)}$ | 13.6 / 18.6 |
| $1/r^5$ | $-\frac{\delta a}{R}\left(-3+\frac{2a}{R}+\frac{R^2}{a^2}\right)\left(-3+\frac{3a}{R}-\frac{a^2}{R^2}+\frac{R}{a}\right)^{-1}$ | $-\frac{\delta a}{a}$ | 12.5 / 17.1 |
| $1/r^6$ | $-\frac{\delta a}{R}\left(-4+\frac{3a}{R}+\frac{R^3}{a^3}\right)\left(-3+\frac{4a}{R}-\frac{3a^2}{2R^2}+\frac{R^2}{2a^2}\right)^{-1}$ | $-\frac{2\delta a}{a}$ | 10.5 / 14.1 |
| $1/r^7$ | $-\frac{\delta a}{R}\left(-5+\frac{4a}{R}+\frac{R^4}{a^4}\right)\left(-\frac{10}{3}+\frac{5a}{R}-2\frac{a^2}{R^2}+\frac{R^3}{3a^3}\right)^{-1}$ | $-3\frac{\delta a}{a}$ | 8.7 / 11.2 |
| $1/r^8$ | $-\frac{\delta a}{R}\left(-6+\frac{5a}{R}+\frac{R^5}{a^5}\right)\left(-\frac{15}{4}+6\frac{a}{R}-\frac{5a^2}{2R^2}+\frac{R^4}{4a^4}\right)^{-1}$ | $-4\frac{\delta a}{a}$ | 7.4 / 9.1 |
| $1/r^9$ | $-\frac{\delta a}{R}\left(-7+6\frac{a}{R}+\frac{R^6}{a^6}\right)\left(-\frac{21}{5}+7\frac{a}{R}-3\frac{a^2}{R^2}+\frac{R^5}{5a^5}\right)^{-1}$ | $-5\frac{\delta a}{a}$ | 6.6 / 7.8 |
| $1/r^{12}$ | $-\frac{\delta a}{R}\left(-10+9\frac{a}{R}+\frac{R^9}{a^9}\right)\left(-\frac{45}{8}+10\frac{a}{R}-\frac{9a^2}{2R^2}+\frac{R^8}{8a^8}\right)^{-1}$ | $-8\frac{\delta a}{a}$ | 5.3 / 5.9 |